# GPU ACCELERATION OF IMAGE CONVOLUTION USING SPATIALLY-VARYING KERNEL

*Steven Hartung\*, Hemant Shukla†, J. Patrick Miller‡ and Carlton Pennypacker§*

\*Centre for Astronomy, James Cook University, Townsville, Australia
†Lawrence Berkeley National Laboratory, Berkeley, CA, USA
‡Department of Mathematics, Hardin-Simmons University, Abilene, TX, USA
§Space Sciences Lab, University of California at Berkeley, CA, USA

## ABSTRACT

Image subtraction in astronomy is a tool for transient object discovery such as asteroids, extra-solar planets and supernovae. To match point spread functions (PSFs) between images of the same field taken at different times a convolution technique is used. Particularly suitable for large-scale images is a computationally intensive spatially-varying kernel. The underlying algorithm is inherently massively parallel due to unique kernel generation at every pixel location. The spatially-varying kernel cannot be efficiently computed through the Convolution Theorem, and thus does not lend itself to acceleration by Fast Fourier Transform (FFT). This work presents results of accelerated implementation of the spatially-varying kernel image convolution in multi-cores with OpenMP and graphic processing units (GPUs). Typical speedups over ANSI-C were a factor of 50 and a factor of 1000 over the initial IDL implementation, demonstrating that the techniques are a practical and high impact path to terabyte-per-night image pipelines and petascale processing.

## 1. INTRODUCTION

The frontiers of astronomy research are expanding into highly sensitive, larger sky-coverage, and multi-band, state-of-the-art telescopes such as Sloan Digital Sky Survey (SDSS) [1], Panoramic Survey Telescope & Rapid Response System (Pan-STARRS) [2], and the proposed Large Synoptic Survey Telescope (LSST) [3]. This trend is rapidly forcing the field into a data deluge on the order of terabyte data rates per night of operations, and petascale data over mission lifetimes. In the case of Pan-STARRS and LSST, nearly every exposure shall be analyzed for transient objects.

To address the scalability and performance required by large data volumes, the current data acquisition, processing and analyses algorithms require review, and in some cases, rewrite. Several efforts are underway to attain the needed high-performance computing by exploiting the emerging hardware availability, and development software support, of massively parallel many-core and accelerator architectures.

In collaboration with one such effort spearheaded at UC Berkeley and Lawrence Berkeley National Laboratory (LBNL), titled Infrastructure for Astrophysics Applications Computing (ISAAC), this work explored a non-traditional and high impact spatially-varying convolution algorithm for performance enhancements using commodity graphics processing units (GPUs). The project is titled IP$^2$ (ISAAC Petascale Image Processing) under which several critical algorithms will be researched and developed for higher performance and scalability on a wide range of architectures.

For ground-based telescopes, atmospheric conditions and optics fluctuate at varying degrees leading to temporally invariant transfer functions known as the point spread function (PSF). In the case of comparing two images, such as for detecting any spatial or brightness changes, it becomes critical to first normalize the images. The technique discussed below convolves one image with a spatially-varying convolution kernel until the first image matches the second image to the degree specified. This process becomes increasingly complex due to pixel and sub-pixel scale image offsets and rotations that redistribute the PSF in unpredictable ways [4]. The sequential implementation of variants of the algorithm available in IDL (Interactive Data Language) and ANSI-C are slow and render the process impractical for larger data volumes. For time-critical astronomical events that demand immediate follow-up such bottlenecks can severely affect the scientific discovery.

Encouraged by the performance enhancements of parallel algorithms on GPUs, the project carefully down-selected the spatially-varying convolution for implementation on a GPU architecture.

## 2. THE PROBLEM

In astronomical images, the fields primarily remain unchanged over time. Therefore many changes are detectable through comparison by image subtraction. The challenge, however, is equalizing the images to counteract optical system differences before any subtraction operation can be performed.

The technique to equalize the images by convolution as discussed in [5] uses a convolution kernel, PSF, to smear the sharper image to match the quality of the second image. In the case of a field with no variable objects, a reference image $R$ convolved with an ideal kernel $K$ should equal another image $I$. The difference image $D$ between the convolved reference and the second image must then yield a constant field (to the level of the background noise).

$$(R \otimes K) - I = D \quad (1)$$

Any significant deviation from the constant value in the difference image D, positive or negative, identifies the position of variable objects in the image field.

In order to determine an appropriate kernel $K$, a statistical technique, named Optimal Image Subtraction (OIS), to match the PSF between images was developed by Alard and Lupton as discussed in [6]. In practice, isolated stars in the reference image serve as the source for PSF determination. The OIS technique is a least-squares minimization method for Equation(1), where the kernel $K$ can be decomposed into a superposition of weighted basis functions. For large fields of view the PSF can vary across the image. Therefore, in practice a single kernel would not always perform well across the entire image. In this spatially-varying convolution, the kernel is allowed to vary across the image with the changing distortions across the image such that,

$$K_{x,y} = \sum_i \left(a_n(x,y)\right)_i B_i \quad (2)$$

where $B_i$ is a basis function, scaled by a bivariate polynomial $a_n$. Where $a_n(x,y)$ is a polynomial of order $n$. In this case the kernel is now unique for every pixel location $(x,y)$. The zeroth order polynomial covers changes in the width of the PSF, the first order corrects for lateral translation, and the second order addresses rotational translation. This study focuses entirely on the $n = 2$ second order polynomial implementation, see Equation (3).

$$a_{x,y,B} = a_{00} + a_{01}y + a_{02}y^2 + a_{11}xy + a_{10}x + a_{20}x^2 \quad (3)$$

A rigorous derivation of spatially-varying convolution technique is described by Alard in [7] with additional details explained by Miller et al in [8].

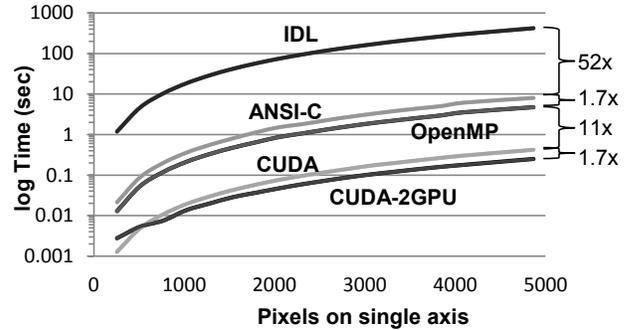

Figure 1. Performance of the DFB spatially-varying kernel convolutions for a range of image sizes using various platforms.

Two types of basis functions are preferred, the Gaussian function basis (GFB) [6] and the Dirac delta function basis (DFB) [8]. The GFB is inherently symmetric and can produce a less than optimal kernel in cases of asymmetric PSF. The DFB can match any profile, but it can become more computationally expensive as it requires a basis function polynomial for every pixel of the kernel. This study focused on the more computationally intensive DFB.

Benchmarking of the optimized IDL implementation [9] of image subtraction code with the computationally intensive DFB reveals that a significant amount of time, more than 96%, is spent in the convolution section of the code.

For comparison, a sequential ANSI-C version of OIS with the GFB, called ISIS [6, 7], was also profiled and revealed similar behavior. In a GFB subtraction, 87% of the computation was devoted to the spatially-varying convolution.

The convolution theorem allows for an acceleration of convolutions by performing highly efficient convolution in the Fourier domain using Fast Fourier Transform (FFT) [10]. However, the spatially-varying convolution cannot take advantage of the convolution theorem and is calculated instead in image space.

These benchmarking results, combined with the inapplicability of the FFT, provide a clear impetus to implement an efficient version of the spatially-varying convolution algorithm on parallel architectures.

## 3. COMPUTATIONAL REQUIREMENTS

This study is focused on testing the efficacy of leveraging parallel architectures for performance enhancement to eliminate (or reduce) bottlenecks in the legacy code instead of porting the entire application. At this stage the IDL code is still used to generate the coefficients for the bivariate polynomials for the DFB technique. The computationally expensive DFB is then convolved for various images sizes of $256 \times 256$ to $4846 \times 4868$.

As an example of computational burden, a $L \times L$ GFB kernel might be well defined in three basis functions and three associated bivariate polynomials, while the DFB for the same $L \times L$ kernel requires $L^2$ basis functions (one for each pixel of the kernel) with $L^2$ associated polynomials. For the DFB, including the final pixel subtraction, the computation load is $20L^2+1$ floating point operations per pixel, which for a realistic $7 \times 7$ kernel example is 981 *FLOP/pixel*.

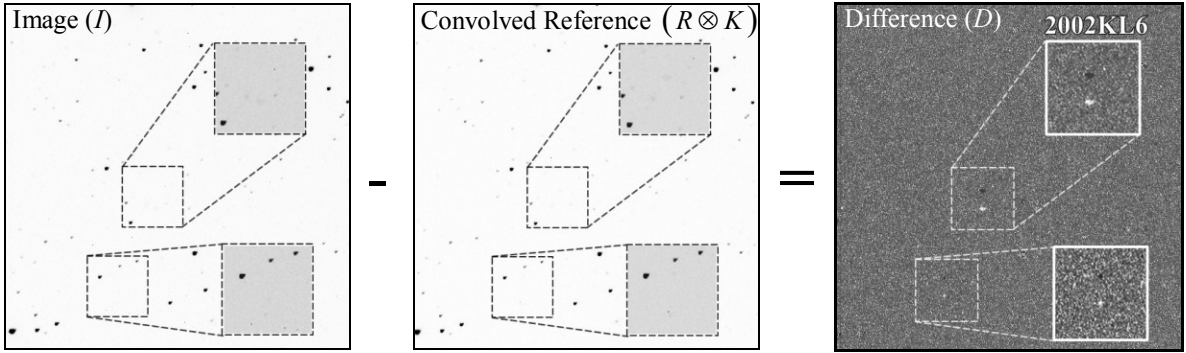

Figure 2. On the left are two negative images of the same star field taken approximately 15 minutes apart. Image Credit: NEAT, courtesy NASA/JPL-Caltech. The zoomed regions are mid-tone enhanced to bring out faint objects. On the right is the DFB $2^{nd}$-order spatially-varying convolution image subtraction result of the two images with moving objects detected.

For a real case wide field mosaic camera (e.g. Pan-STARRS) with tile sizes of $4846 \times 4868$ pixels, the DFB estimation translates to 23.1GFLOP/*tile* or 1.39TFLOP/image for the 60 tile camera. The GFB with N basis functions and an $L \times L$ kernel is $NL^2+2L^2+18N+1$ floating point operations per pixel. For a 3 basis GFB and a $7 \times 7$ kernel, that is 300 *FLOP/pixel*, for similar wide field images with the $4846 \times 4868$ pixel tile, that amounts to 7.08 *GFLOP/tile* or 425 *GFLOP/image* for the 60 tile camera.

These are all ideal best cases, and do not include the practicalities of setting up and controlling loop variables etc. In practical application with existing single threaded code, computation is often reduced by calculating a new kernel only at certain pixel intervals, at the expense of kernel precision. This study calculated the best available kernel at every pixel.

## 4. IMPLEMENTATION - OPENMP, CUDA

The IDL code for the DFB spatially-varying convolution was first ported to a new single threaded ANSI-C implementation that was suitable for subsequent parallelization under OpenMP and CUDA. This provided a basis to examine both small scale and larger scale parallelization. The code developed for this study uses the two original images and a list of bivariate polynomial coefficients generated in IDL as inputs. The output consists of an image subtraction result and timing data.

The independent single pixel based kernel determination allows for allocating thread resources to each pixel and thus leveraging inherent layout of optimal threads in thread-blocks for the GPU without any thread-block boundary effects. The data naturally lends to memory coalesced data access as adjacent pixels provide the convolution halo for each other.

Basic optimizations were incorporated in the CUDA code, which resulted in significant speedups. Through experimental tuning, the images were tiled in $32 \times 32$ sized chunks for each thread block, which align with CUDA thread *Warps*, significantly increasing the performance. Image data used asynchronous transfers between pinned host memory and GPU device memory. The polynomial calculation was unrolled (also unrolled in the ANSI-C and OpenMP implementation). While it is possible to unroll the convolution itself, tests showed the gains to be minimal and compiling for specific kernel sizes is required.

The algorithm as described by Alard [7] and Miller et al [8] defines the kernel changes over the entire image space in a compact parameterized form of polynomial coefficients, a development which is very useful for GPU optimization. The coefficients for the pixel specific kernel generation were stored in GPU constant memory, thus allowing all convolution threads to access a single copy of the coefficients at near register speeds.

Given the large size of the convolution halo around each pixel, the limited shared memory space constrains its usability for halo pixels. Relying on the GPU data cache pipeline hardware appears to outperform identifying and moving the halo pixels into shared memory in experiments thus far. The large per-pixel computational load, combined with the zero-overhead thread switching, provides adequate latency masking for image data accesses from the GPU global memory. This may change as new GPU hardware generations add to the available shared memory.

CUDA streams and two identical GPUs were employed in parallel on a single node in the final test.

## 5. RESULTS

Figure (1) shows that the pixel-unique kernel algorithm as taken from legacy IDL code and rewritten in C provides significant speedup. What is evident from the plots is that parallelization due to multiple cores leads to tremendous improvements in performance. As can be seen at the lower left in Figure 1, multi-GPU is not recommended for small images since times actually increased for smaller images as compared to a single GPU.

The following hardware and software were employed in this study: CPU - Athlon II X3, 3.1GHz triple-core (64-bit); GPU - NVIDIA GTX465(x2), Fermi, 352-core, 1GB DDR5; CUDA 3.2; gcc 4.4.3-1ubuntu1; GCC OpenMP 4.4.3-4ubuntu5; and IDL 8.1.

Strictly from a domain science perspective, starting with widely used IDL code, 3 orders of magnitude speedup has been achieved by using very low-cost hardware.

### 5.1 Example Application

The DFB technique was applied to two images from the NEAT project [11]. Images were obtained from the SkyMorph image server provided by NASA/GSFC. Moving objects in image subtraction appear as a pair of light and dark objects as the subtraction identifies them as missing in one image and appearing in the other. In the test case, the subtraction in Figure (2) yields an otherwise almost invisible near-Earth object (NEO asteroid 2002KL6) that has spatially moved in the temporally separated images. A second even fainter object also appears moving in a similar orbit, this second object was not anticipated by the authors and at the time of this paper it remains unidentified by the authors. The object 2002KL6 is at magnitude 22.7 in this image, just above the background level. The grain in the resulting image is the background noise enhanced by the contrast change from the reduced dynamic range after subtraction.

## 6. CONCLUSIONS

The spatially-varying convolution method, as used in astronomical image subtraction, is made practical for large data by the application of parallel processing. Multicore or GPU architectures provide for easily attainable and effective scaling, but massively parallel low-cost GPU is a particularly effective approach for this problem. With very modest GPU hardware, the performance of the second order DFB spatially-varying convolution improved by three orders over the publically available IDL implementation. Future work will explore additional optimizations in GPU tuning, multi-node scaling using MPI, as well as the limits of effective parallelism. Additional work will improve performance on other functions of the OIS method now that some of them take longer than the spatially-varying kernel convolution. A complete pipeline-ready tool will be made available.

This study has helped in isolating a significant bottleneck that has been encapsulated for implementation in different programming models. Thereby, lending more flexibility and preventing the code from vendor-locking. In the future efforts, an OpenCL version may allow for additional flexibility in hardware selection.

In addition, the authors would like to emphasize that despite the fact that the algorithm is developed here as an astronomy application, it can be adapted in other domains that analyzes changes in before and after images, for example medical imaging, geo-sciences, etc.

## 7. ACKNOWLEDGMENT

HS would like to acknowledge the ICCS project ISAAC funded through the NSF grant award #0961044 under PI Dr. Horst Simon for supporting this work. SH and JPM would like to thank Dr. William Burgett and Dr. Larry Denneau for their support of Pan-STARRS (University of Hawaii) test images for this work. SH and JPM would like to thank Tomas Vorobjov for his assistance in attempting to identify the detected objects in the example NEAT images. Portions of this work were supported by a James Cook University research grant, GRS 6648.93001.0208.